\documentclass[aps,PRB,twocolumn,superscriptaddress,preprintnumbers]{revtex4-2}
\usepackage[T1]{fontenc}
\usepackage{times}
\usepackage{graphicx,color}
\usepackage{amsfonts,amsmath,amssymb,amsbsy}
\usepackage[colorlinks=true, linkcolor=blue, citecolor=blue, urlcolor=blue]{hyperref}
\usepackage{float}

\begin{document}

\title{Bulk photovoltaic effects in altermagnets}
\author{Motohiko Ezawa}
\affiliation{Department of Applied Physics, The University of Tokyo, 7-3-1 Hongo, Tokyo
113-8656, Japan}

\begin{abstract}
The bulk photovoltaic effect is a photocurrent generation from alternating
electric field, which is a promising candidate for future efficient solar
cell technology. It is the second-order optical current, which is the
injection current or the shift current. We focus on the direct current
generation. By employing a simple two-band model of the $d$-wave altermagnet
coupled with the Rashba interaction, we show that the linearly polarized
light can generate the injection and shift currents when the N\'{e}el vector
points to an in-plane direction. The magnitude of the injection current is
almost constant over a wide range of the frequency $\omega $ of the applied
light provided it is smaller than a certain critical frequency $\omega _{%
\text{c}}$ and larger than the bulk gap energy $\varepsilon _{\text{gap}}$, $%
\varepsilon _{\text{gap}}<\hbar \omega <\hbar \omega _{\text{c}}$. Hence,
the use of the injection current is quite efficient for solar cell
technology because any photon whose energy is within this range can be
equally utilized.
\end{abstract}

\date{\today }
\maketitle

\textbf{Introduction:} Nonlinear optical responses are fascinating, upon
which there are intensive researche. When we apply an alternating electric
field $E(\omega )$ with the frequency $\omega $, there are two types of the
second-order responses. One is the second-harmonic generation proportional
to $2\omega $. The other is a direct current (dc) generation. The latter is
important in the context of photovoltaic effects. Recently, bulk
photovoltaic effects\cite{Beli,Kraut,Baltz,Ave,Sipe,Frid} including the
injection current\cite%
{Sipe,Ave,JuanNC,Juan,AhnInject,AhnNP,WatanabeInject,Dai} and the shift
current\cite%
{Young,Young2,Kraut,Baltz,Ave,Sipe,Juan,AhnInject,MorimotoScAd,Kim,Barik,AhnNP,WatanabeInject,Dai,Yoshida}
attract much attention because they provide more efficient photovoltaic
effects than the usual one based on the $p$-$n$ junction for the application
to solar cell devices. Indeed, only the photon having the same energy $\hbar
\omega $ as the band gap is transformed to a current in the $p$-$n$\
junction. On the other hand, the second-order current is generated for the
photon with the energy larger than the band gap, which is the injection
current or the shift current. There are several experimental observation of
bulk photovoltaic effects\cite{Braun,Nakamura,Sotome,Oste,Hatada}.

Altermagnets constitute one of the most active fields of condensed matter
physics\cite{SmejRev,SmejX,SmejX2}. It has zero net magnetization as in the
case of an ordinary antiferromagnet, but breaks time-reversal symmetry and
has a momentum dependent energy spectrum\cite%
{Naka,Gonza,NakaB,Bose,NakaRev,AhnAlter,Hayami,SmejRev,SmejX,SmejX2}. These
features open a way to future ultrahigh density spintronic memories with
ultrafast switching in the order of ps\cite{Takagi} because there is no
stray field owing to the zero net magnetization. There are some studies on
nonlinear responses as well on the altermagnet\cite%
{YFang,EzawaMetricC,EzawaGI}.

\begin{figure}[t]
\centerline{\includegraphics[width=0.48\textwidth]{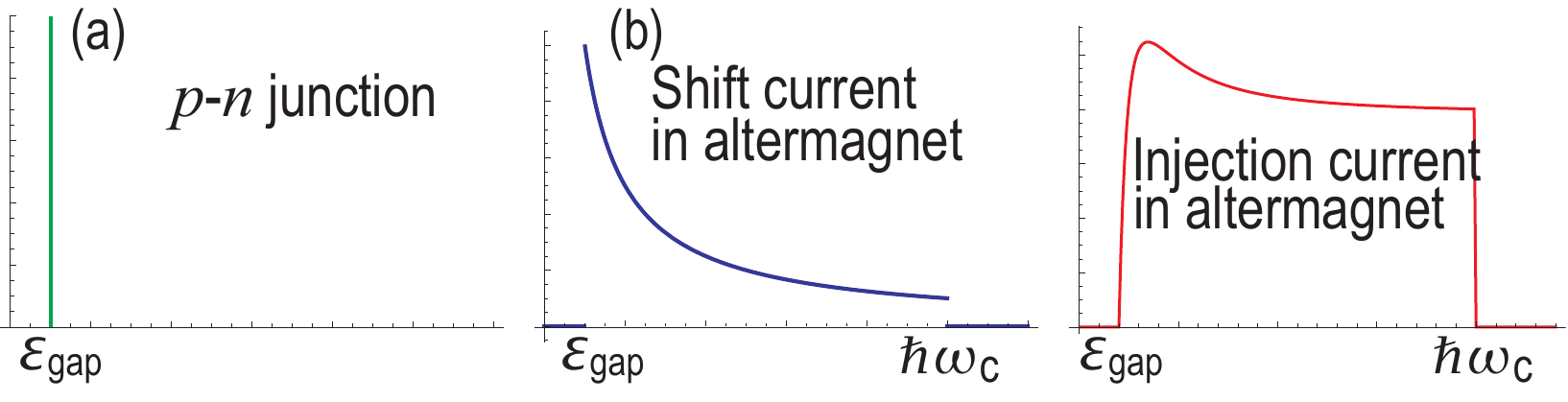}}
\caption{Illustration of photovoltaic current as a function of the energy $%
\hbar \protect\omega $ of the photon. (a) In the case of the $p$-$n$
junction, only a photon with the energy $\hbar \protect\omega =\protect%
\varepsilon _{\text{gap}}$ contributes to the current. (b) A photon with the
energy $\protect\varepsilon _{\text{gap}}<\hbar \protect\omega <\hbar 
\protect\omega _{\text{c}}$ contributes to the shift current but the
efficiency decreases for large energy as $1/\protect\omega $, where $\hbar 
\protect\omega _{\text{c}}$ is a certain critical photon energy. (c) A
photon with the energy $\protect\varepsilon _{\text{gap}}<\hbar \protect%
\omega <\hbar \protect\omega _{\text{c}}$ contributes almost equally to the
injection current. The vertical axis is the magnitude of photovoltaic
current, while the horizontal axis is the energy $\hbar \protect\omega $ of
the applied photon.}
\label{FigIllust}
\end{figure}

In this paper, we study photovoltaic effects in the $d$-wave altermagnet
with the Rashba interaction based on a simple two-band model, where the N%
\'{e}el vector points to an in-plane direction. The dc current generation by
the second-order optical process is studied. We summarize the results in Fig.%
\ref{FigIllust}. The photocurrent generation occurs only when the energy of
a photon is within a certain range $\varepsilon _{\text{gap}}<\hbar \omega
<\hbar \omega _{\text{c}}$, where $\varepsilon _{\text{gap}}$ is the bulk
gap energy and $\omega _{\text{c}}$ is the critical frequency. A photon
contributes to the shift current but the efficiency decreases for large
energy as $1/\omega $ as in Fig.\ref{FigIllust}(b). On the other hand, a
photon contributes almost equally to the injection current as in Fig.\ref%
{FigIllust}(c). Hence, it would be useful to employ the $d$-wave altermagnet
to generate the injection current for future solar cell technology.
Incidentally, in the $p$-$n$\ junction, only a photon whose energy is
identical to the bulk gap $\varepsilon _{\text{gap}}$ contributes to the
current ($\hbar \omega =\varepsilon _{\text{gap}}$) as in Fig.\ref{FigIllust}%
(a).

\textbf{Injection current and shift current:} The current density $j$
induced by the applied electric field $E$ is expanded as%
\begin{equation}
j^{c}=\sigma ^{c;a}E_{a}+\sigma ^{c;ab}E_{a}E_{b}+\cdots .
\end{equation}%
The first term is the linear response and the second term is the
second-order nonlinear response. If we apply an alternating electric field,
the second-order response has a form%
\begin{equation}
j^{c}\left( \omega _{1}+\omega _{2}\right) =\sigma ^{c;ab}\left( \omega
_{1}+\omega _{2};\omega _{1},\omega _{2}\right) E_{a}\left( \omega
_{1}\right) E_{b}\left( \omega _{2}\right) .
\end{equation}%
In this paper, we investigate the dc current generation, 
\begin{equation}
j^{c}\left( 0\right) =\sigma ^{c;ab}\left( 0;\omega ,-\omega \right)
E_{a}\left( \omega \right) E_{b}\left( -\omega \right) .
\end{equation}%
In the following, we use the abbreviation $j^{c}\equiv j^{c}\left( 0\right) $
and $\sigma ^{c;ab}\left( \omega \right) \equiv \sigma ^{c;ab}\left(
0;\omega ,-\omega \right) $.

We apply incident light propagating along the $z$\ direction. We consider
the linear polarized light,%
\begin{equation}
E^{\updownarrow }=E_{0}\left( \cos \phi ,\sin \phi ,0\right) 
\end{equation}%
with\ the polarization $\phi $, and the circularly polarized light,%
\begin{equation}
E^{\circlearrowleft }=\frac{E_{0}}{\sqrt{2}}\left( 1,i,0\right) ,\qquad
E^{\circlearrowright }=\frac{E_{0}}{\sqrt{2}}\left( 1,-i,0\right) .
\end{equation}%
The conductivities are given by\cite{AhnInject,JiangRev}%
\begin{eqnarray}
\sigma ^{c;\updownarrow }\left( \Phi \right)  &=&\text{Re}\sigma ^{c;xx}\cos
^{2}\Phi +\text{Re}\sigma ^{c;yy}\sin ^{2}\Phi   \notag \\
&&+2\text{Re}\sigma ^{c;xy}\cos \Phi \sin \Phi ,  \notag \\
\sigma ^{c;\circlearrowleft } &=&\text{Re}\sigma ^{c;xx}+\text{Re}\sigma
^{c;yy}+2\text{Re}\sigma ^{c;xy},  \notag \\
\sigma ^{c;\circlearrowright } &=&\text{Re}\sigma ^{c;xx}+\text{Re}\sigma
^{c;yy}-2\text{Im}\sigma ^{c;xy}.  \label{polari}
\end{eqnarray}

\begin{figure}[t]
\centerline{\includegraphics[width=0.48\textwidth]{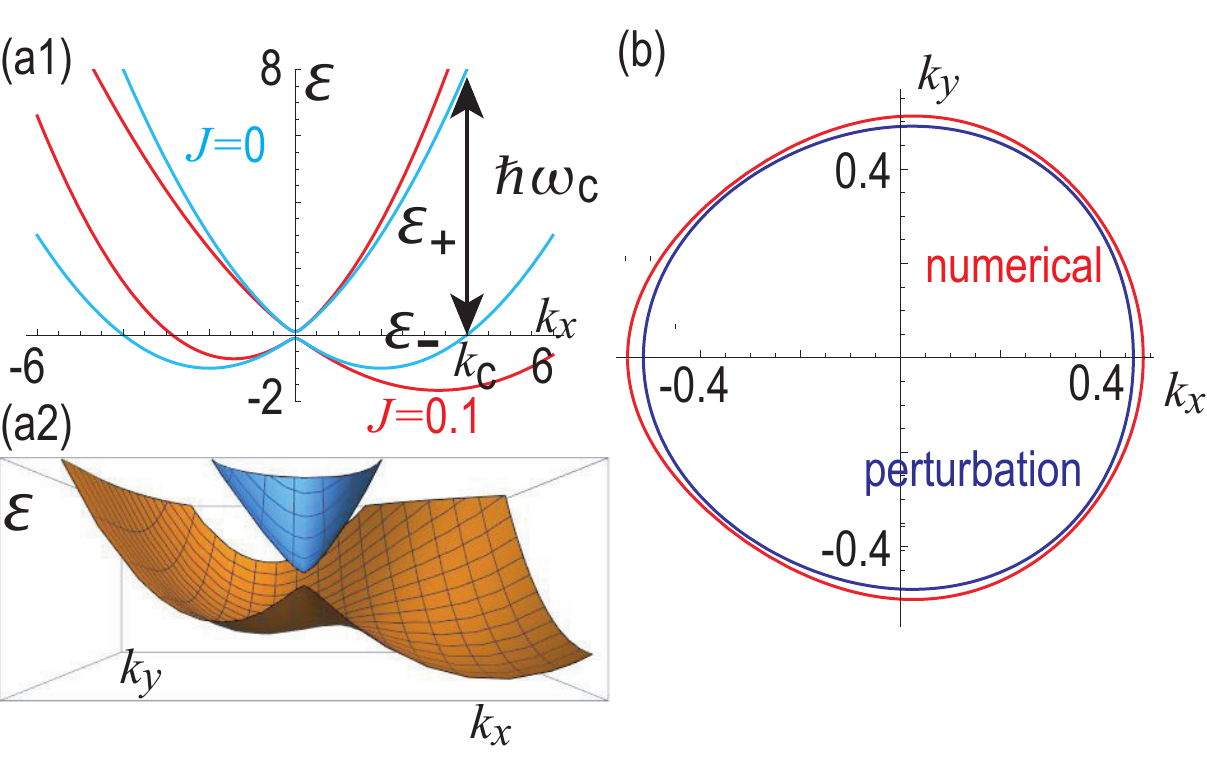}}
\caption{(a1) Energy spectrum $\protect\varepsilon _{\pm }\left( \mathbf{k}%
\right) $. The vertical axis is the energy $\protect\varepsilon $\ in units
of $\protect\varepsilon _{0}$. The horizontal axis is $k_{x}$ in units of $%
k_{0}$. The red and cyan curves represent the energy spectrum in the case of 
$J=0.1\protect\varepsilon _{0}/k_{0}^{2}$ and $J=0$, respectively. (a2)
Bird's eye's view in the case of $J=0.1\protect\varepsilon _{0}/k_{0}^{2}$.
(b) $k\left( \protect\phi \right) $ at $\hbar \protect\omega =0.4\protect%
\varepsilon _{0}$ in the case of $J=0.1\protect\varepsilon _{0}/k_{0}^{2}$.
The red ellipse is the Fermi surface of the original model, while the blue
ellipse is that of the perturbation theory given in Eq.(\protect\ref{kAppro}%
). We have set $B=0.1\protect\varepsilon _{0}$.}
\label{FigBand}
\end{figure}

The injection current is in general given by the formula\cite%
{Sipe,Ave,JuanNC,Juan,AhnInject,AhnNP,WatanabeInject,Okumura,Dai}%
\begin{align}
\sigma _{\text{inject}}^{c;ab}=& -\tau \frac{2\pi e^{3}}{\hbar ^{2}}\int 
\frac{d^{3}k}{\left( 2\pi \right) ^{3}}\sum_{n,m}\left( f_{n}-f_{m}\right)
\Delta _{mn}^{c}  \notag \\
& \times r_{nm}^{b}r_{mn}^{a}\delta \left( \omega _{m}-\omega _{n}-\omega
\right) ,  \label{Inject}
\end{align}%
where $\tau $ is the relaxation time, $a$ is the lattice constant, $%
f_{n}=1/\left( \exp \left( \varepsilon _{n}-\mu \right) +1\right) $ is the
Fermi distribution function for the band $n$, $\mu $ is the chemical
potential, $\varepsilon _{n}$ is the energy of the band $n$, $%
r_{mn}^{a}=\left\langle m\right\vert i\partial _{k_{a}}\left\vert
n\right\rangle $ is the Berry connection and $\Delta
_{mn}^{c}=v_{mm}^{c}-v_{nn}^{c}$ is the interband transition of the velocity 
$v_{mn}^{c}=\frac{1}{\hbar }\left\langle m\right\vert \partial
_{k_{c}}H\left\vert n\right\rangle $. The injection current is nonzero when
the velocity of the energy dispersion is imbalanced between the conduction
and valence bands along the $c$ direction.

On the other hand, the shift current is in general given by the formula\cite%
{Young,Young2,Kraut,Baltz,Ave,Sipe,Juan,AhnInject,MorimotoScAd,Kim,Barik,AhnNP,WatanabeInject,Dai,Yoshida}%
\begin{align}
\sigma _{\text{shift}}^{c;ab}=& -\frac{\pi e^{3}}{\hbar ^{2}}\int \frac{%
d^{3}k}{\left( 2\pi \right) ^{3}}\sum_{n,m}\left( f_{n}-f_{m}\right)
(R_{mn}^{c,a}-R_{nm}^{c,b})  \notag \\
& \times r_{nm}^{b}r_{mn}^{a}\delta \left( \omega _{m}-\omega _{n}-\omega
\right) ,  \label{Shift}
\end{align}%
where $R_{mn}^{c,a}=r_{mm}^{c}-r_{nn}^{c}+i\partial _{k_{c}}\log r_{mn}^{a}$
is the shift vector\cite{Sipe}. The shift vector is gauge invariant although
the Berry connection is not gauge invariant. The shift vector describes the
mean position of the Wannier function. The shift current is nonzero when the
mean positions are different between the conduction and valence bands.

It is known\cite%
{Holder,Semenov,Ogawa,Chan,YZhang,ZSun,Iguchi,WatanabeInject,Dai} that the
injection (shift) current can be generated by the linearly (circularly)
polarized light in magnetic system due to the time-reversal symmetry
breaking.

\textbf{Altermagnet:} We consider a two-dimensional system made of the $d$%
-wave altermagnets with the Rashba interaction, whose Hamiltonian is given by%
\cite{SmejRev,SmejX,SmejX2}%
\begin{align}
H\left( \mathbf{k}\right) =& \frac{\hbar ^{2}\left(
k_{x}^{2}+k_{y}^{2}\right) }{2M}I_{2}+\lambda \left( k_{x}\sigma
_{y}-k_{y}\sigma _{x}\right)  \notag \\
& +J\left( k_{x}^{2}-k_{y}^{2}\right) \mathbf{n}\cdot \mathbf{\sigma }%
+B\sigma _{z},  \label{dModel}
\end{align}%
where $M$ is the effective mass of the free electrons, $I_{2}$\ is the $%
2\times 2$ identity matrix, $\lambda $ is the magnitude of the Rashba
interaction, $J$ is the magnitude of the $d$-wave altermagnetization, and $%
\mathbf{n}$ is the N\'{e}el vector of the $d$-wave altermagnet. We set $%
\mathbf{n}=\left( 0,1,0\right) $. The Rashba interaction is introduced by
placing an altermagnet on the substrate\cite%
{SmejRev,SmejX,SmejX2,Zu2023,Gho,Li2023,EzawaAlter,EzawaMetricC,EzawaGI}.
The last term $B\sigma _{z}$ is introduced by magnetization, or by applying
an external magnetic field, or by the Edelstein effect due to the Rashba
splitting under in-plane electric field.

The $d$-wave magnet in two dimensions is realized in organic materials\cite%
{Naka}, perovskite materials\cite{NakaB}, and twisted magnetic Van der Waals
bilayers\cite{YLiu}. The $d$-wave altermagnet in three dimensions is
realized in RuO$_{2}$\cite{Ahn,SmeRuO,Tsch,Fed,Lin}, Mn$_{5}$Si$_{3}$\cite%
{Leiv} and FeSb$_{2}$\cite{Mazin}.

The Hamiltonian (\ref{dModel}) preserves inversion symmetry in the absence
of the Rashba interaction. The Rashba interaction breaks inversion symmetry,
while the altermagnet term breaks time-reversal symmetry. Hence, the system
breaks both inversion symmetry and time-reversal symmetry. The inversion
symmetry breaking is necessary for the presence of the second-order optical
responses. We assume $\left\vert J\right\vert <\hbar ^{2}/\left( 2M\right) $
so that the parabolic dispersion is positive for large $k=|k|$.

\begin{figure}[t]
\centerline{\includegraphics[width=0.48\textwidth]{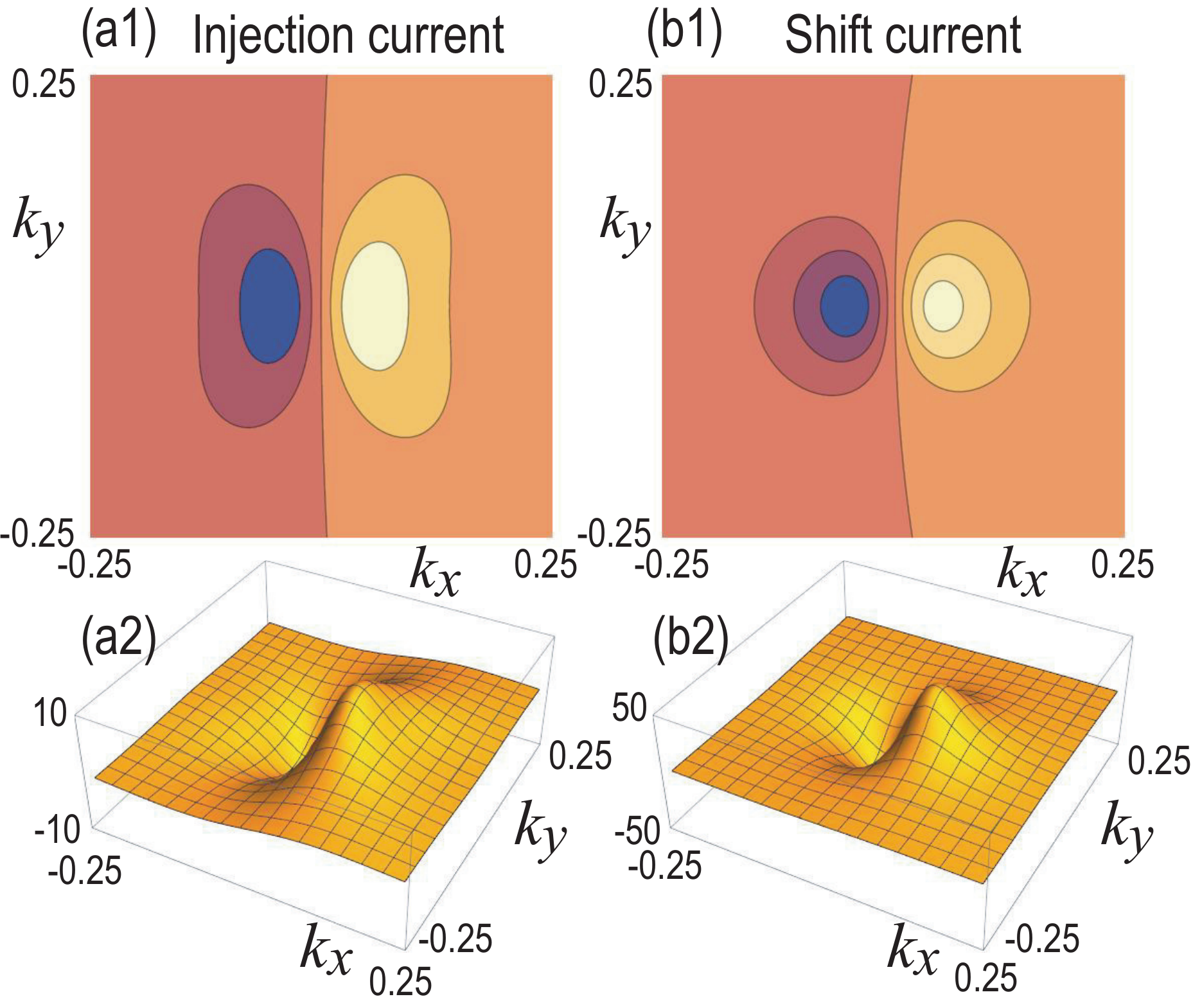}}
\caption{(a1), (a2) $\Delta _{+-}^{x}\left\vert r_{-+}^{x}\right\vert ^{2}$\
in the $k_{x}$-$k_{y}$\ plane for the injection current. (b1), (b2) Re$%
\left( R_{+-}^{x,x}-R_{-+}^{x,y}\right) r_{-+}^{y}r_{+-}^{x}$\ in the $k_{x}$%
-$k_{y}$\ plane for the shift current. (a1), (b1) Contour plot. (a2), (b2)
Bird's eye's view. We have set $J=0.1\protect\varepsilon _{0}/k_{0}^{2}$\
and $B=0.1\protect\varepsilon _{0}$. }
\label{FigDensity}
\end{figure}

\textbf{Photovoltaic effects:} The Hamiltonian (\ref{dModel}) is of the form,%
\begin{equation}
H=h_{0}\left( \mathbf{k}\right) I_{2}+\sum_{j=x,y,z}h_{j}\left( \mathbf{k}%
\right) \sigma _{j},
\end{equation}%
describing a two-band system. The energy spectrum consists of $\varepsilon
_{\pm }\left( \mathbf{k}\right) =h_{0}\left( \mathbf{k}\right) \pm
\varepsilon \left( \mathbf{k}\right) $ with $\varepsilon \left( \mathbf{k}%
\right) =\sqrt{\sum_{j=x,y,z}h_{j}^{2}}$.

The energy spectrum is illustrated in Fig.\ref{FigBand}(a), where there is a
Dirac cone at the $\Gamma $ point ($k_{x}=k_{y}=0$) with the bulk band gap $%
\varepsilon _{\text{gap}}=2\left\vert B\right\vert $. We set the chemical
potential zero ($\mu =0$). We now examine the condition imposed on the
frequency $\omega $\ for the photocurrent generation to occur. The optical
transition occurs from the occupied valence band ($f_{-}=1$) to the
unoccupied conduction band ($f_{+}=0$) at the energy $\varepsilon
_{+}-\varepsilon _{-}=2\varepsilon =\hbar \omega $. Thus, we may set $%
f_{-}-f_{+}=1$ in the injection current (\ref{Inject}) and the shift current
(\ref{Shift}). Furthermore, by examining the energy spectrum in Fig.\ref%
{FigBand}(a), the optical transition is found to occur only when the
frequency $\omega $\ of a photon is within the range $\varepsilon _{\text{gap%
}}<\hbar \omega <\hbar \omega _{\text{c}}$, where the critical frequency $%
\omega _{\text{c}}$ is determined as $\hbar \omega _{\text{c}}=2\varepsilon (%
\mathbf{k}_{\text{c}})$ with the use of $\mathbf{k}_{\text{c}}$ which is the
solution of $\varepsilon _{-}(\mathbf{k}_{\text{c}})=0$ for the conduction
band.

We study the injection and shift currents under the linearly and circularly\
polarized light. We may set $m=+$ and $n=-$ in Eqs.(\ref{Inject}) and Eq.(%
\ref{Shift}) for a two-band system. By introducing the polar coordinate of
the momentum, $k_{x}=k\cos \phi $, $k_{y}=k\sin \phi $, and by setting $%
\varepsilon (\mathbf{k})\equiv \varepsilon (k,\phi )$, it is straightforward
to derive the formula%
\begin{align}
& \sigma _{\text{inject}}^{x;xx}  \notag \\
=& -\tau \frac{2\pi e^{3}}{\hbar ^{2}W}\int \frac{kdkd\phi }{\left( 2\pi
\right) ^{2}}\left( f_{-}-f_{+}\right) \Delta
_{+-}^{x}r_{-+}^{x}r_{+-}^{x}\delta \left( \frac{2\varepsilon }{\hbar }%
-\omega \right)   \notag \\
=& -\tau \frac{2\pi e^{3}}{\left( 2\pi \right) ^{2}\hbar ^{2}W}\int k\left. 
\frac{\Delta _{+-}^{x}\left\vert r_{-+}^{x}\right\vert ^{2}}{2\left\vert
\partial _{k}\varepsilon \right\vert /\hbar }\right\vert _{k=k_{\omega
}\left( \phi \right) }\hspace{-5mm}d\phi   \label{InjectB}
\end{align}%
for the injection current (\ref{Inject}), and 
\begin{align}
\sigma _{\text{shift}}^{x;xy}& =-\frac{\pi e^{3}}{\hbar ^{2}W}\int \frac{%
kdkd\phi }{\left( 2\pi \right) ^{2}}\left( f_{-}-f_{+}\right) \left(
R_{+-}^{x,x}-R_{-+}^{x,y}\right) r_{-+}^{y}r_{+-}^{x}  \notag \\
& \;\;\;\;\;\;\times \delta \left( \frac{2\varepsilon }{\hbar }-\omega
\right)   \notag \\
& =-\frac{\pi e^{3}}{\left( 2\pi \right) ^{2}\hbar ^{2}W}\int k\left. \frac{%
\left( R_{+-}^{x,x}-R_{-+}^{x,y}\right) r_{-+}^{y}r_{+-}^{x}}{2\left\vert
\partial _{k}\varepsilon \right\vert /\hbar }\right\vert _{k=k_{\omega
}\left( \phi \right) }\hspace{-5mm}d\phi   \label{ShiftB}
\end{align}%
for the shift current (\ref{Shift}), where $W$ is the width of the sample
given by $W^{-1}=\left( 2\pi \right) ^{-1}\int dk_{z}$, and $k_{\omega
}\left( \phi \right) $ is given by solving $2\varepsilon \left( k_{\omega
}\left( \phi \right) ,\phi \right) =\hbar \omega $.

\begin{figure}[t]
\centerline{\includegraphics[width=0.48\textwidth]{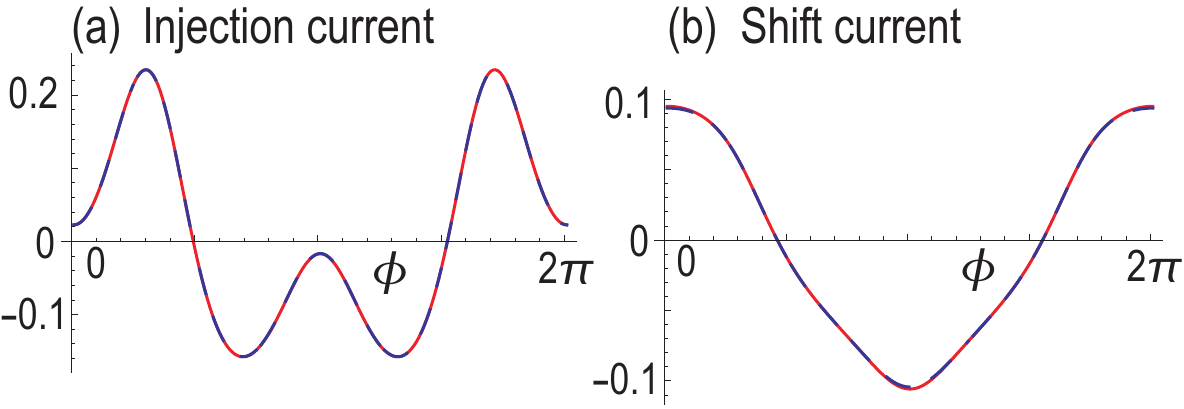}}
\caption{(a) $\Delta _{+-}^{x}\left\vert r_{-+}^{x}\right\vert ^{2}$ as a
function of $\protect\phi $ for the injection current. (b) Re$\left(
R_{+-}^{x,x}-R_{-+}^{x,y}\right) r_{-+}^{y}r_{+-}^{x}$ as a function of $%
\protect\phi $ for the shift current. The blue dashed curves are the
numerical results without using the perturbation theory, while the red
curves are the analytical results based on the perturbation theory. The
agreement is remarkable. We have set $\hbar \protect\omega =0.4\protect%
\varepsilon _{0}$, $J=0.1\protect\varepsilon _{0}/k_{0}^{2}$\ and $B=0.1%
\protect\varepsilon _{0}$.}
\label{FigCut}
\end{figure}

It is notable that the injection current and the shift current do not depend
on the effective mass $M$. This is because the Berry connection $r_{mn}^{a}$%
, the interband transition of the velocity $\Delta _{mn}^{c}$\ and the shift
vector $R_{mn}^{c,a}$\ are solely determined by $h_{j}$\ with $j=x,y,z$.

We determine the critical frequency $\omega _{\text{c}}$\ exactly in the
case of $J=0$. The condition $\varepsilon _{-}(\mathbf{k}_{\text{c}})\equiv
\varepsilon _{-}(k_{\text{c}},\phi _{\text{c}})=0$\ gives%
\begin{equation}
k_{\text{c}}=\sqrt{2}\sqrt{M^{2}\lambda ^{2}+M\sqrt{M^{2}\lambda ^{4}+B^{2}}}%
.
\end{equation}%
Then, the critical energy is obtained as%
\begin{equation}
\hbar \omega _{\text{c}}=2\varepsilon \left( k_{\text{c}}\right) =2\sqrt{%
\lambda ^{2}k_{\text{c}}^{2}+B^{2}}.
\end{equation}%
The effect of $J$\ to $\omega _{\text{c}}$\ is found to be tiny comparing
with the bulk band gap $2\left\vert B\right\vert $. We derive various
formulas valid up to the first order in $J/\left( W\lambda \right) $ in what
follows.

\begin{figure}[t]
\centerline{\includegraphics[width=0.48\textwidth]{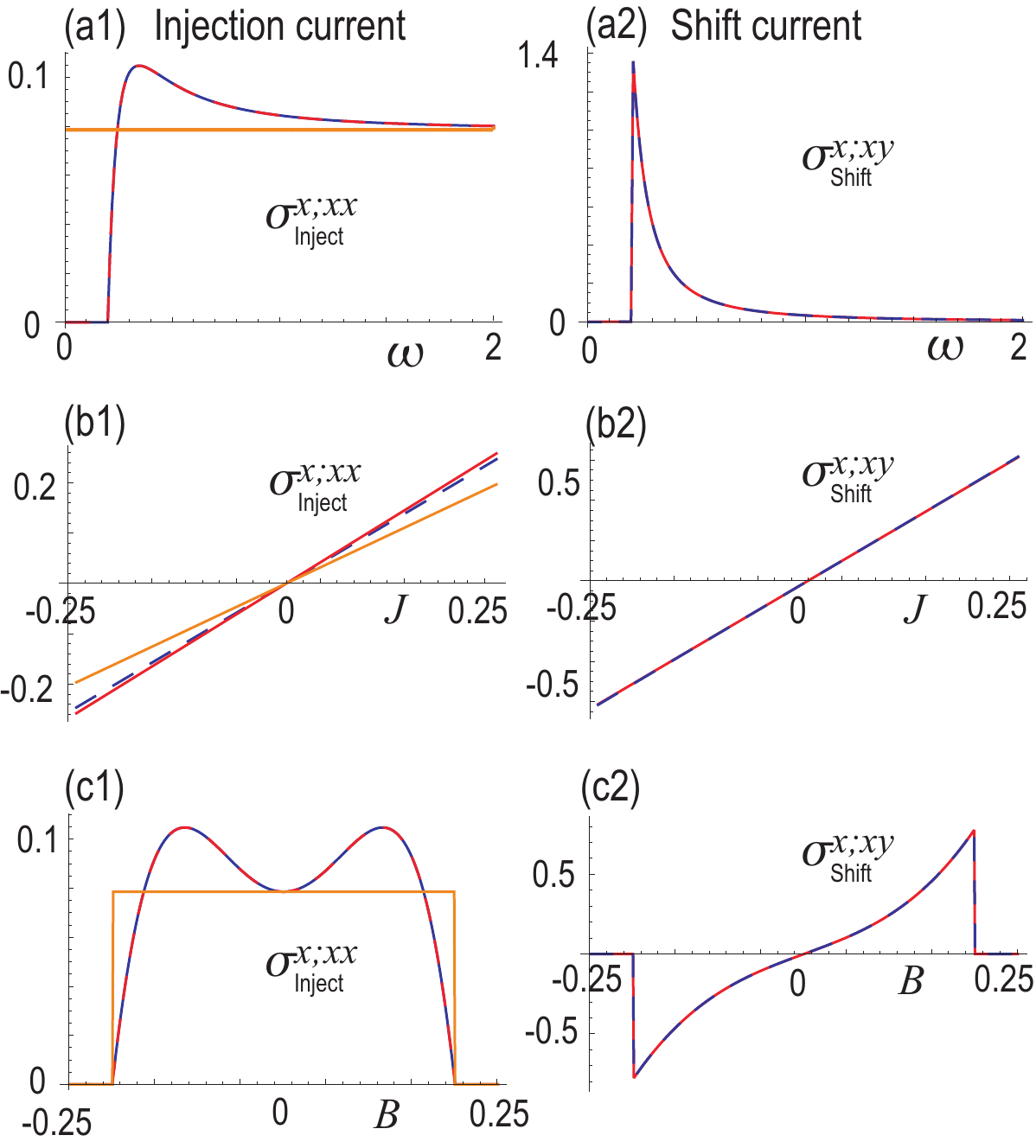}}
\caption{(a1), (a2) Current as a function of $\hbar \protect\omega /\protect%
\varepsilon _{0}$. We have set $J=0.1\protect\varepsilon _{0}/k_{0}^{2}$\
and $B=0.1\protect\varepsilon _{0}$. (b1), (b2) Current as a function of $%
\frac{J}{W\protect\lambda }$. We have set $B=0.1\protect\varepsilon _{0}$\
and $\hbar \protect\omega =0.4\protect\varepsilon _{0}$. (c1), (c2) Current
as a function of $B/\protect\varepsilon _{0}$. We have set $J=0.1\protect%
\varepsilon _{0}/k_{0}^{2}$\ and $\hbar \protect\omega =0.4\protect%
\varepsilon _{0}$. (a1), (b1), (c1) Injection current $\protect\sigma _{%
\text{inject}}^{x;xx}$ in units of $-\protect\tau \frac{e^{3}}{8\hbar ^{2}}%
\frac{J}{W\protect\lambda }$. (a2), (b2), (c2) The shift current $\protect%
\sigma _{\text{shift}}^{x;xy}$ in units of $\frac{e^{3}}{4\hbar ^{2}}\frac{J%
}{W\protect\lambda }$. The blue dashed curves are the numerical results
without using the perturbation theory, while the red curves are the
analytical results based on the perturbation theory. The agreement is
remarkable. Orange lines are the result for $B\ll \hbar \protect\omega $
given in Eq.(\protect\ref{limit}). The critical energy is $\hbar \protect%
\omega _{\text{c}}\simeq 8\protect\varepsilon _{0}$. }
\label{FigInject}
\end{figure}

We show the energy spectrum $\varepsilon _{\pm }\left( \mathbf{k}\right) $
along the $k_{x}$ axis in Fig.\ref{FigBand}(a). It is asymmetric along the $%
k_{x}$ direction for $J\neq 0$. It enables the emergence of the injection
current and the shift current. The Fermi surface is shown as a red ellipse
in Fig.\ref{FigBand}(b). Up to the first order in $J/\left( W\lambda \right) 
$, the energy of the Hamiltonian (\ref{dModel}) is given by%
\begin{equation}
\varepsilon (k,\phi )=\frac{\hbar ^{2}k^{2}}{2M}+\sqrt{\lambda
^{2}k^{2}+B^{2}}+\frac{Jk^{3}\left( \cos \phi +\cos 3\phi \right) }{\sqrt{%
\lambda ^{2}k^{2}+B^{2}}}.
\end{equation}%
To calculate Eqs.(\ref{InjectB}) and (\ref{ShiftB}), it is necessary to
solve $2\varepsilon (k_{\omega }\left( \phi \right) ,\phi )=\hbar \omega $.
Solving it we obtain%
\begin{equation}
k_{\omega }\left( \phi \right) =\frac{\sqrt{\left( \hbar \omega \right)
^{2}-4B^{2}}}{2\lambda }-\frac{J\left( \left( \hbar \omega \right)
^{2}-4B^{2}\right) \left( \cos \phi +\cos 3\phi \right) }{8\lambda ^{3}}.
\label{kAppro}
\end{equation}%
It is shown as a cyan ellipse in Fig.\ref{FigBand}(b). It well reproduces
the numerical result without using the perturbation theory shown in red
ellipse.

\textbf{Injection current:} We study the injection current by applying
linearly polarized light. The integrand $\Delta _{+-}^{x}\left\vert
r_{-+}^{x}\right\vert ^{2}$\ in Eq.(\ref{Inject}) is shown in the $k_{x}$-$%
k_{y}$\ plane in Fig.\ref{FigDensity}(a1) and (a2). The integration is done
on the ellipse $k_{\omega }\left( \phi \right) $\ given in Eq.(\ref{kAppro}%
), which is shown in Fig.\ref{FigCut}(a). The injection current Eq.(\ref%
{InjectB}) is obtained up to the first order in $J/\left( W\lambda \right) $
as%
\begin{equation}
\sigma _{\text{inject}}^{x;xx}=-\tau \frac{e^{3}}{8\hbar ^{2}}\frac{J}{%
W\lambda }\left( 1+8\left( \frac{B}{\hbar \omega }\right) ^{2}-48\left( 
\frac{B}{\hbar \omega }\right) ^{4}\right) .  \label{InjectXX}
\end{equation}%
When the applied energy $\omega $ is much larger than the band gap $%
\varepsilon _{\text{gap}}=2\left\vert B\right\vert $, the injection current
is independent of $B$ and has a simple form%
\begin{equation}
\sigma _{\text{inject}}^{x;xx}=-\tau \frac{e^{3}}{8\hbar ^{2}}\frac{J}{%
W\lambda },  \label{limit}
\end{equation}%
which is the illustration in Fig.\ref{FigIllust}(c). The injection current
is shown as a function of $\omega $ in Fig.\ref{FigInject}(a1), as a
function of $J$ in Fig.\ref{FigInject}(b1), and as a function of $B$ in Fig.%
\ref{FigInject}(c1). The numerical result and the perturbation result well
agree one to another. The injection current is almost independent of the
applied frequency $\omega $.

Similary, we have 
\begin{equation}
\sigma _{\text{inject}}^{x;yy}=-\tau \frac{e^{3}}{8\hbar ^{2}}\frac{J}{%
W\lambda }\left( 1-2\left( \frac{B}{\hbar \omega }\right) ^{2}+4\left( \frac{%
B}{\hbar \omega }\right) ^{4}\right) ,
\end{equation}%
and 
\begin{equation}
\sigma _{\text{inject}}^{x;xy}=-\tau \frac{e^{3}}{8\hbar ^{2}}\frac{iJ\pi }{%
\lambda }\frac{B}{\hbar \omega }\left( 1-4\left( \frac{B}{\hbar \omega }%
\right) ^{2}\right) .
\end{equation}%
Then, the injection currents induced by the circularly polarized light (\ref%
{polari}) are given by%
\begin{equation}
\sigma _{\text{inject}}^{x;\circlearrowleft }=-\tau \frac{e^{3}}{8\hbar ^{2}}%
\frac{J}{W\lambda }\left( 2+2B\right) +o\left( B^{3}\right) ,
\label{InjectXYA}
\end{equation}%
and 
\begin{equation}
\sigma _{\text{inject}}^{x;\circlearrowright }=-\tau \frac{e^{3}}{8\hbar ^{2}%
}\frac{J}{W\lambda }\left( 2-2B\right) +o\left( B^{3}\right) .
\label{InjectXYB}
\end{equation}

\textbf{Shift current:} Next, we study the shift current. The real part of $%
\left( R_{+-}^{x,x}-R_{-+}^{x,y}\right) r_{-+}^{y}r_{+-}^{x}$\ in Eq.(\ref%
{Shift}) is shown\ in the $k_{x}$-$k_{y}$\ plane in Fig.\ref{FigDensity}(b1)
and (b2). The integration is done on the ellipse $k_{\omega }\left( \phi
\right) $\ given in Eq.(\ref{kAppro}), which is shown in Fig.\ref{FigCut}%
(b). The transverse conductivity $\sigma _{\text{shift}}^{x;xy}$ Eq.(\ref%
{ShiftB}) is obtained up to the first order in $J/\left( W\lambda \right) $
as

\begin{equation}
\sigma _{\text{shift}}^{x;xy}=\frac{1}{\omega }\frac{e^{3}}{4\hbar ^{2}}%
\frac{J}{W\lambda }\frac{B}{\hbar \omega }\left( 1+4\left( \frac{B}{\hbar
\omega }\right) ^{2}\right) ,  \label{ShiftXY}
\end{equation}%
which is the illustration in Fig.\ref{FigIllust}(b). It vanishes for the
gapless system with $B=0$. The current is generated proportional to $%
1/\omega $. It is shown as a function of $\omega $ in Fig.\ref{FigInject}%
(a2), as a function of $J$\ in Fig.\ref{FigInject}(b2) and as a function of $%
B$\ in Fig.\ref{FigInject}(c2). It is induced by the linearly polarized
light with $\phi =\pi /4$. On the other hand, the shift current is not
generated by applying the circularly polarized light because the transverse
conductivity $\sigma _{\text{shift}}^{x;xy}$\ is real\cite{Sipe}, where the
imaginary part of the transverse conductivity is necessary\cite{AhnInject}.

On the other hand, explicit calculations show that the shift current is zero 
$\sigma _{\text{shift}}^{x;xx}=\sigma _{\text{shift}}^{x;yy}=0$, when the
linearly polarized light is applied.\ Hence, the shift current is not
generated by applying the circularly polarized light,%
\begin{equation}
\sigma _{\text{shift}}^{x;\circlearrowleft }=\sigma _{\text{shift}%
}^{x;\circlearrowright }=0.
\end{equation}

\textbf{Discussion:} Bulk photovoltaic effects are useful for future solar
cell technology because they produce the direct current from alternating
electric field. Solar light has a continuous frequency spectrum. In the
viewpoint of applications, the injection current has greater merits than the
shift current. Indeed, all photons within a wide range of frequency
contribute almost equally to the injection current. On the other hand,
although the shift current is generated from all photons within the same
range, the magnitude is proportional to $1/\omega $. Namely, the
contribution from photons with high frequency $\omega $ is small.
Furthermore, the injection current has a finite contribution even for $B=0$.
It means that $B$ can be infinitesimally small because it is only necessary
for making a finite gap so that the Berry connection is well defined. The
geomagnetism may be enough for small $B$. On the other hand, the shift
current is proportional to $B$, where it is necessary to introduce large $B$%
. It is done by attaching a ferromagnet to the sample. However, it degrades
the transparency to inject a light into the sample. Finally, the injection
current is proportional to the relaxation time $\tau $. Then, the injection
current is enhanced for a clean sample. On the other hand, the shift current
is independent of $\tau $, where the shift current cannot be enhanced even
for a clean sample.

In this work we made the perturbation in the parameter $J/\left( W\lambda
\right) $, whose validity we discuss. A typical value of the Rashba
interaction is $\lambda =0.33$eV\r{A}\ at the Au(111) surface\cite{La}. A
typical value\cite{YLi} of $J$\ is $J=30$meV$\times a^{2}$\ with the lattice
constant $a$, where $a$\ is of the order of $10$\r{A}. Hence, we have $%
J/\left( W\lambda \right) \sim a/W\ll 1$ and the perturbation is justified,
where the width $W$\ is typically $W>100a$. 

We make a comment on the terminology in Ref.\cite{Wang}, where\ the
injection current (\ref{Inject}) is decomposed into the normal and magnetic
injection currents, and a similar decomposition for the shift current (\ref%
{Shift}). Here, only the normal ones emerge in the system with time-reversal
symmetry, while both the normal and magnetic ones may emerge in the system
without time-reversal symmetry. The present system (\ref{dModel}) has no
time-reversal symmetry. The injection current (\ref{InjectXX}) induced by
the linear polarized light is identical to the magnetic injection current%
\cite{Wang,YZhang}, while the injection currents (\ref{InjectXYA}) and (\ref%
{InjectXYB}) induced by circularly polarized light are checked to be the
normal injection currents and there is no magnetic injection current. On the
other hand, the shift current induced by linearly polarized light (\ref%
{ShiftXY}) is the normal shift current and there is no shift current induced
by the circularly polarized light.

The main result of this work is that the injection current is almost
independent for $\varepsilon _{\text{gap}}<\hbar \omega <\hbar \omega _{%
\text{c}}$, which is derived based on a simple two-band model. This result
may be modified in actual experiments due to two reasons\cite{JuanNC}.
First, the complex band structures of real materials cannot be accurately
captured by this simple two-band model. Second, the effects of electron-hole
interactions or many-body interactions may change the spectrum of the
injection current. It would be interesting to make an experimental study of a
system made of $d$-wave altermagnets with the Rashba interaction, where
various energy can be harvested over a wide range of $\omega $\ via the
injection current.

The author is very much grateful to T. Morimoto and N. Nagaosa for helpful
discussions on the subject. This work is supported by CREST, JST (Grants No.
JPMJCR20T2) and Grants-in-Aid for Scientific Research from MEXT KAKENHI
(Grant No. 23H00171).


\end{document}